# Functionally Fractal Urban Networks:

# Geospatial Co-location and Homogeneity of Infrastructure

Christopher Klinkhamer[1, *], Elisabeth Krueger[1,2], Xianyuan Zhan[1],

Frank Blumensaat[3], Satish Ukkusuri[1], and P. Suresh C. Rao[1,4]

**1**/ Lyles School of Civil Engineering, Purdue University, West Lafayette, IN 47907, USA

**2**/ Helmholtz Centre for Environmental Research - UFZ, Leipzig, Germany

**3**/ Institute of Environmental Engineering, ETH, Zurich, Switzerland

**4**/ Agronomy Department, Purdue University, West Lafayette, IN 47907, USA

**Abstract**

Just as natural river networks are known to be globally self-similar, recent research has shown that human-built urban networks, such as road networks, are also functionally self-similar, and have fractal topology with power-law node-degree distributions ($p(k) = a\,k^{-\gamma}$). Here we show, for the first time, that other urban infrastructure networks (sanitary and storm-water sewers), which sustain flows of critical services for urban citizens, also show scale-free functional topologies. For roads and drainage networks, we compared functional topological metrics, derived from high-resolution data (70,000 nodes) for a large US city providing services to about 900,000 citizens over an area of about 1,000 km². For the whole city and for different sized subnets, we also examined these networks in terms of geospatial co-location (roads and sewers). Our analyses reveal functional topological homogeneity among all the subnets within the city, in spite of differences in several urban attributes. The functional topologies of all subnets of both infrastructure types resemble power-law distributions, with tails becoming increasingly power-law as the subnet area increases. Our findings hold implications for assessing the vulnerability of these critical infrastructure networks to cascading shocks based on spatial interdependency, and for improved design and maintenance of urban infrastructure networks.

**Introduction**

Cities are important case studies of human-dominated ecosystems, where dynamics of flows of resources such energy, food, and water support human communities, and are examples of coupled relationships between humans and nature [1]. Feedbacks between humans and technical systems within the city have cascading impacts over much larger spatiotemporal scales beyond the urban boundaries [2, 3]. It is known that physical assets in cities are fractal, and so too are road networks that constrain the geometry of the urban fabric [4-6]. We posit that several infrastructure networks that support critical urban services are also located in close physical proximity to one another, and as such might also share functional topological features.

Studying urban infrastructure networks that support flows of critical services to communities distributed over large expanses is important for understanding urban dynamics, and for examining impacts on ecosystems beyond cities. Urban networks can be studied from either a structural perspective (e.g., evolution of road networks [4, 7] over time) or from a functional perspective (e.g., dynamics of flows on a network [6, 8]). Here, we conceive urban infrastructure as complex networks, and apply graph theory concepts and network analyses to examine similarities and differences in topologies of above- and below-ground infrastructure networks, and compare them to our recent analyses [9] of urban sewer networks and their natural analogs, river networks; both networks serving the same function: efficient drainage of landscapes.

We begin by providing an analysis of two urban infrastructure networks, roads and sewers, in a large US city (~900,000 residents within the study area of nearly 1,000 $km^2$) [10], based on high-resolution spatial data (~70,000 nodes), to reveal for the first time, the similarities in functional topology of these networks, and the spatial topological homogeneity of different sizes of infrastructure subnets within the city with evident differences in urban structure (city center vs. suburbs) and other attributes.

Prior studies [11] of small water distribution networks have found them to be sparse and uninformative without accounting for the heterogeneity in importance of certain key features (reservoirs, tanks, pumps, etc.) within the networks, and suggest that functional properties are of major consequence in the analysis of water distribution networks and may be accounted for in the form of weighted networks [12]. Rather than analyzing weighted networks, functional aspects, related to flows, of the network can be implemented by considering the dual representation, where segments, defined by some attribute (such as street name or curvature; or pipe size), are instead considered as nodes and each intersection or junction an edge. We utilize a dual-mapping method, known as Hierarchical Intersection Continuity Negotiation (HICN) [13] to examine road and sewer networks. Previous studies have successfully utilized this and similar methods to produce dual representations to show the universality of functional topological attributes in road networks [5, 8, 13-16].

For reasons of practicality, reliability, and eminent domain, the geospatial location of many infrastructure networks, including road networks and other underground water networks (urban drainage; sanitary sewers; potable water distribution) are expected to be well correlated. Practical considerations, such as intense competition for space within urban areas, the ability to easily locate, access, and maintain existing infrastructure, and governmental control of land through eminent domain, further reinforce that urban infrastructure networks be spatially co-located [17-19].

Here, we first examine the functional topologies of road and drainage networks in the study area focusing on their node-degree probability density distributions (pdfs). In the next section, we explore the spatial variability of the functional topology of multiple sizes of subnets within the city. Finally, the geospatial co-location of roads and below-ground drainage networks and their topological properties are evaluated. We close with implications of our findings to the design and maintenance of urban infrastructure and community resilience.

**Results**

    A) **Study Area and Infrastructure Network Topology**

The topography of the studied urban area is flat, with moderate slopes and rolling hills, experiencing mid-continental climate with four distinct seasons. Peak rain occurs from May-July (mean annual rainfall: 16-19 mm/day) [10]. Urban drainage infrastructure for the case study city is a mix of public and privately owned utilities, consisting of both combined and separated sanitary and storm water sewers. High-resolution urban drainage and road network records are maintained by the city municipal GIS office, and were provided for our analyses presented here.

Dual representations of both road and sewer networks for the study area were generated, following the HICN method described by Masucci et al. [13] (details are given in Methods section). Fitting dual-mapped (HICN) node-degree probability density distributions (*pdf*s) for urban infrastructure network data is confounded by several constraints related to availability of sparse data over a limited data range, especially when these distributions are to be approximated as power-law distributions, in particular the "finite-size" effects (see details in methods). For these reasons, our analyses recognize these challenges by estimating power-law distributions with both frontal (to account for minimum node-degree and network resolution) and distal truncation (to acknowledge the finite-size effect) over the limited data range. Thus, double-truncation (Double Power-law) was examined similar to the method used by Massucci et al [13] and following the power-law fitting guidelines proposed by Clauset et al. [20], and further refined by Corral and Deluca [21].

For both randomly generated subnets, and for the city as a whole, the node-degree *pdf*s of all dual-mapped road networks larger than 5 km$^2$, and dual-mapped sewer networks larger than 100 km$^2$, reveal two distributions. These node-degree *pdf*s were approximated as double power-law with greater than 90 % of tests (2,500 repetitions per subnet) failing to reject the null-hypothesis of a power-law distribution. The slopes of the entirety of the road and sewer networks were found to be approximately $\gamma_{r1}$ = 2.5 and $\gamma_{r2}$ = 3.1 for roads, and $\gamma_{s1}$ = = 2.9 and $\gamma_{s2}$ = = 4.1 for sewers. The results for the upper distributions ($\gamma_{x1}$) are in accordance with numerous other complex network studies indicating a narrow range of $\gamma$-values of about -2 to -3 [22-24] and are consistent among subnets of different sizes and location with minimal variability, especially when a subnet area greater than 20 km$^2$ is considered. The results for the tails of the *pdf*s ($\gamma_{x2}$) are both more variable and steeper than the upper trunk segments, and are influenced by the finite-size effect, and seem to gradually approach the slope of the upper segment as the subnet size is increased.

Confirming this convergence and comparing the slopes ($\gamma$) for different types and sizes of networks, along with the degree of truncation, should be a topic of future research and would allow for drawing inferences about likely differences in network topology. In addition, the emergence of power-law *pdf*s with increasing network size in road and water networks in diverse cities would also be useful in uncovering consistent patterns in urban infrastructure network topology, and thus provide an indication of the underlying generating mechanisms for the infrastructure networks.

### B) Spatial Homogeneity of Functional Network Topology

Here, self-similar and scaling properties of these infrastructure networks were investigated by means of producing randomly distributed, nested subnets of 1.25, 2.5, 5, and 7.5 km radius, and analyzing four common topological metrics: node-degree distribution, average node-degree, clustering coefficient, and network density. In addition to the spatial co-location of roads and drainage pipes, a better understanding of the scaling and self-similar properties exhibited by these infrastructure networks would be a valuable resource for city planners and decision makers seeking to predict the location of water pipes in a city. If the water and drainage networks can be shown to be *functionally self-similar* at various scales, only a small area of sufficient size may be used to represent the city as a whole with reasonable certainty, rather than having to consider the entire city (a process requiring a great deal of data and computer processing time for large cities).

Beyond a threshold population of ~5% (~50,000 persons; 20 $km^2$ area) of the total, the average node-degree of both roads and pipes was found to be highly homoscedastic, with little variability throughout the city (Figure 3-A). This suggests that, spatially or with population variability, there is an overall homogeneity in terms of how these sewer and road networks are connected, i.e., the ratio of edges to nodes remained more or less constant among the subnets. The values of the average node-degree also provide insight to the physical structure of these networks. In the limits, a large average node-degree (near four), such as those found for roads, indicates a network with a highly looped structure, whereas a small average node-degree (near two), such as those found for pipes, indicates a more tree-like structure with fewer loops. These differences in structure provide important insights into the design principles based on functions (e.g., flow directionality) and performance reliability/redundancy demands of these networks.

Roads and sewers serve inherently different purposes. Sewers are primarily concerned with collection of inputs from multiple, spatially distributed sources, aggregation into larger pipes before reaching a single (the wastewater treatment plant) or few (combined sewage outfalls) destination points. For such a network, with converging flows towards an outlet, a tree-like structure is often the most efficient mode of transport, as is well known to be the case for natural river networks [25, 26]. Roads, on the other hand, must take multiple inputs (origins, drivers) and allow them to reach *any* possible destination within the city. While the optimal solution for a single origin or driver would likely be a spanning-tree network reaching all destinations, in order to accommodate multiple origins and multiple drivers needing to reach multiple destinations, overlapping trees must be created and the resulting irregular lattice-like road structure is familiar to most modern cities.

In addition to structural considerations, the average node-degree and corresponding tree-like or looped structure hints towards the perceived importance of the two types of infrastructures. Access to reliable transportation is well correlated with increased economic output, and an improved standard of living [27]. As such, it is critical that the transportation system does not fail. The grid-like structure of urban road networks allows for redundancy and alternate routes in the event of failure (congestion) at a given node. Flows in sewer networks are primarily gravity-driven, and failures are relatively infrequent (flooding during large return-period storm events, clogging, pipe collapses) or moderate (minor roadway flooding or subsidence) as long as major components (large node-degree) are segregated from other critical infrastructure or services (major roads, electric components, hospitals, etc.) [28]. In addition, cost efficiency dictates that loops should occur infrequently given that sewer pipes should be installed along the shortest path possible at a minimum depth, and minimum capacity to meet service demands.

Based on the clustering coefficients, both road and sewer networks in the study area were found to be spatially homogenous in functional topology, again with a homoscedastic distribution after a threshold population of approximately 50,000 persons (see Figure 3-B), reinforcing the idea that these networks are similarly connected at all scales, i.e., self-similar topology. The small values of the clustering coefficients indicate that neighbors in these networks are not well connected and the network does not exhibit small world characteristics [29].

Overall, the clustering coefficient for the sewer networks was nearly double that of the road networks, indicating a greater tendency for roads to cluster together in cliques as compared to sewer networks in the same study area. This finding is in accordance with standard principles of sewer design. Being predominantly gravity-driven and converging-flow networks, sewer networks are often, though not always, organized to follow natural watersheds with only minimal connections between neighboring watersheds and a priority towards efficiently removing water from a given area. Roads and drivers on the other hand are less constrained by topography and may freely move between watersheds at multiple locations with a priority towards ease of mobility.

The average node-degree and clustering coefficients together show that road and sewer infrastructure networks in *this city* tend to be highly spatially homogenous, and reveal these networks to be self-similar at multiple scales throughout the city. At each spatial scale, the node-degree distribution, average node-degree, and clustering coefficient **all** fall within narrow ranges with minimal variability, indicating that the ways in which these networks connect remain constant across spatial scale. Given this self-similarity, one would expect network density to decrease as the total number of possible edges increases approaching the limit at zero as city (i.e., entire network) size is increased towards infinity. In the smallest subnet these networks contain an average of ~1,000 nodes and edges while the city as a whole contains upwards of 70,000 nodes and edges. Indeed, we observed the network density of both types of infrastructure networks to scale with population following a truncated power-law slope, providing further affirmation of the self-similar, scale-free nature of these networks (Figure 3-C). Analyses of infrastructure network data from several cities is needed to examine if the spatial homogeneity pattern is a common feature.

The hierarchical nature of each infrastructure network was compared by plotting the average node-degree vs. the corresponding clustering coefficient (Figure 3-D). In a perfectly hierarchical network, wherein identical subcomponents are connected to each other perfectly and repeatedly, the slope of such a plot should result in a power-law distribution with negative slope and $\gamma = 1$ [30, 31]. Here, our results indicate that the plot of average node-degree vs clustering coefficient for the study area road network results in a power-law distribution, with $\gamma = 0.9$, indicating a highly regular and hierarchical network. Such a plot for sewers, in contrast, has a positive slope, owing to the gravity driven nature of these systems and a tendency to connect to any larger pipe capable of accommodating the designed maximum discharge. These connections are often irregular and produce a non-hierarchical, loop-less network structure with various sized and shaped components optimally connecting wherever is most convenient and cost-effective.

### C) Geospatial Co-location of Infrastructure Networks

These topological analyses however do not take into account spatial information, such as where important, high node-degree pipes or roads are located. Thus, the spatial orientation of roads and sanitary sewers were analyzed for the study area. Starting with a roadway centerline shapefile (provided by the municipal GIS office), buffers of 1 m increments (1 to 15 m range) were applied to the centerline and intersecting sanitary sewer lines were clipped out. In the United States the design standard for lane width is 3.6 m [32], meaning that the maximum buffering distance in this study is equivalent to the width of a two-lane road plus an equivalent area of right-of-way (land obtained through eminent domain) surrounding the road. The mean width of the actual right-of-way was found to be equivalent to an 11 m buffer on both sides of the centerline.

For the studied urban area, the total length of sewers located under the buffered area increased from ~18 % of the total length of the sewer system at a 1 m buffering distance, to ~74 % at the 15 m buffer (Figure 4-A) ~66 % of the total length of the sewer system was found to fall within the boundaries of the mean right-of-way distance. At a 15 m buffer distance, ~49 % of the length of the road network can be expected to have a sewer pipe beneath, and ~66 % of all roads can be expected to have a sanitary pipe beneath at least some portion of its length. Indeed, much of the length of the sewer network for this city can be expected to be found in close physical proximity to roads. These findings are in accordance with those for the city of Innsbruck, Austria [33, 34], and support the assumption that pipes are under roads [33-35].

In addition to co-location, correlation between the *size* of a pipe or *class* of a road and the underlying sewer pipes is essential for assessing the vulnerability of co-located infrastructure networks to cascading failures. Analysis of the size distribution of sewer pipes co-located with roads shows no correlation either positive or negative, for the co-location of large diameter pipes with roads (Figure 4-B). In fact, contrary to what might be expected, large diameter pipes (>250 cm diameter) are found to be almost entirely located under roads, while small (<75 cm diameter) and medium (75-250 cm diameter) pipes exhibit a lesser degree of co-location with roads. Over time, leaking sewer pipes can lead to soil subsidence undermining the integrity of overlying road segments and eventually cause disruptions ranging from small potholes to complete collapse of a road segment. In addition, high traffic and heavy loads on major roadways may increase the deterioration rate of subsurface pipes. For these reasons, among others, city planners would be wise to avoid spatial co-location of major pipes and roads.

When network metrics are considered (i.e., betweenness centrality or node degree as proxy measures for the importance of each pipe or road segment in a network) of co-located segments are considered, a clear pattern emerges (Figure-5: A and B). A clustering of points in the bottom left corner of Figure 5-A indicates a preference for the co-location of low node-degree roads and pipes. While the upper left and bottom right hand quadrants indicate a preference for high node-degree pipes to be located under low node-degree roads, and for high node-degree roads to be located above low node-degree pipes respectively. An absence of points in the upper right hand corner suggests an aversion to the co-location of high node-degree pipes with high node-degree roads. In Figure 5-B the same pattern emerges when the colocation of sewers and roads based on betweenness centrality is considered. Again, an absence of points in the upper right quadrant suggests an aversion to the co-location of high centrality roads and sewers. Tendencies to separate the location of high node-degree and high centrality roads and sewer pipes are also visually evident (Figure 6). Both high-node-degree as well as high-centrality roads and sewers (magnitude represented by line thickness) are seen to be largely separate from each other with only minimal overlap.

**Implications**

Previous topological analyses of natural and human-built networks [4-7, 16, 23-26, 36] offer important guidance to our analyses of urban infrastructure networks presented here. While engineering analyses of the design and functions of urban infrastructure are well understood [32, 37], exploration of urban water network topologies, from a graph theory perspective, has received limited attention to date. Records detailing the location and attributes of above-ground infrastructure networks, such as roads, are often readily available and reliable, allowing for monitoring and modeling of features such as traffic flow [28, 33-35, 38-41]. However, risk and resilience analyses of critical urban infrastructure, those considering the performance, and mitigation of consequences resulting from system failures, require detailed data regarding the specific layout of above-ground and underground networks, as well as additional information on various physical attributes (e.g., traffic volume for roads; pipe size, flows, and connectivity for sewer networks; etc.), which are frequently not available. Thus, comparing topological similarities of roads and rivers with sewer networks will help in developing a general understanding of urban networks.

The similarities in functional topologies of two infrastructure networks (roads and drainage), and that of river networks, hold meaningful implications for their overall reliability under stress. Previous studies [22, 42-44] have shown that scale-free networks (those with power-law node-degree *pdf*s) are highly resilient to random failures but vulnerable to failures of high-degree nodes, which can quickly lead to network-wide discontinuity and system failure. The spatial co-location of urban infrastructure networks, while practical, also introduces the possibility of undesirable consequences. For instance, a leaky pipe can result in soil subsidence and potholes in the road above or a burst water main might cause collapse of road segments, local flooding, and bring traffic to a halt even though the road itself might not have suffered any physical harm or degradation. Co-location of high node-degree or high centrality road and sewer features could exacerbate the issues of interdependence in the event of failure of hubs in one or both networks leading to the potential for cascading failures across networks. Such considerations should be accounted for and minimized during the design process, and included as a component when assessing interdependence of infrastructure networks is evaluated.

The spatial and network analyses performed for one large US metropolitan area in this study indicate that while the co-location of high node-degree and high centrality roads and sewers is generally avoided, in the case of the study city, outliers do exist. These co-located large node-degree features may be thought of as potential points of failure (loss of service; erosion of resilience), and can be used as a criterion for allocating resources for maintenance and service loss prevention.

Beyond the potential for cascading failures, the principles of self-similarity present in the underground sewer lines should allow modelers to develop more realistic hydraulic models given limited data, or setting rules for generating semi-virtual networks [33-35]. By setting rules for the clustering of sewer networks, degree of co-location, avoidance of co-located large node-degree roads and pipes, and overall tree-like structure, modelers can eliminate unrealistic designs that do not conform to known structural and functional features. A highly looped sewer network for instance would likely be unrealistic and cost-ineffective. While such a design would be preferable for high-value, critical, or variably loaded infrastructure networks such as roads.

**Acknowledgements**

The authors wish to acknowledge the city municipal GIS office for their assistance accessing and interpreting the data used in this study. This research was supported by NSF Award Number 1441188 (Collaborative Research-- RIPS Type 2: Resilience Simulation for

Water, Power and Road Networks). Two authors (CK, XZ) were funded by the NSF grant, while EK was supported by the Helmholtz Center for Environmental Research, Leipzig, Germany, and by a Graduate Fellowship from the Purdue Climate Change Research Center. Additional financial support for the last author (PSCR) was provided by the Lee A. Rieth Endowment in the Lyles School of Civil Engineering, Purdue University.

**Competing Financial Interests**

The authors report no competing financial interests.

**Author Contributions**

CK, EK, and PSCR planned the research, and wrote the main text. XZ, SU, and FB provided conceptual support, while EK and XZ provided data analysis. All authors reviewed and contributed to the paper.

**Methods**

**A) Data Acquisition and Preparation:**

Infrastructure data were obtained from the municipal GIS office and consisted of ESRI shapefiles indicating the spatial location of roads, sanitary sewer lines, and associated attributes (such as speed limit, street name, and road class for roads and pipe diameter, and construction material for the sanitary sewer lines). Starting with the existing shapefiles, z-measures were removed so that the networks could be approximated as planar graphs and existing segments were snapped to the nearest closest feature at a threshold of .25m in order to ensure network connectivity.

We analyze the spatial variability and scaling properties of the water and road networks in the study area. For this study, different size subnets were created. Randomly distributed and nested subnets were extracted from each network layer. Subnet creation consisted of selecting 25 random points throughout the city based on the location of manholes, and then running a clip process to extract circular selections surrounding each random point at buffered radii of 1.25, 2.5, 5, and 7.5 km. The population for each subnet was determined using data from the 2010 US census [10].

**B) Network Extraction and Dual Mapping**

For these analyses, the dual representation of the networks was utilized following the method of Hierarchical Intersection Continuity Negotiation (HICN) described by Masucci et al [13]. Generally urban infrastructure networks are analyzed using the primal representation where intersections or junctions are considered as nodes and the connecting segments as edges [23, 24]. This process has been utilized in several studies of urban infrastructure networks [11, 12, 36], and can be useful for determining geometric properties of the network, but by treating network as homogenous (all nodes and links are of equal value and identical function) these types of analyses often obscures functional properties.

**C) Network Topological Metrics**

Here, we investigate four measures of network topology:

1. **Node-degree Distribution [$p(k)$]:** The node-degree *pdf* describes the overall connectivity of the network, i.e. the relative distribution of highly connected nodes to poorly connected nodes. In real-world networks, the node degree distribution will often resemble a power-law distribution, frequently with either frontal or distal truncation or both. Constraints to generating statistically robust estimation of power-law parameters using observational data include: (1) the "finite-size effect" (urban agglomerations are $\leq 10^3$ km$^2$; (2) total number of primal and dual-mapped nodes $\leq 10^4$; and (3) dual-mapped node-degrees $\leq 10^2$). Thus, the available network data, even at high resolution, as in this study, do not cover the multiple orders of magnitude needed to test for "pure" power-law *pdf*s. These challenges become more apparent when data for different sized sub-nets are analyzed for comparison or when network growth over time is examined.

    Here, we have fitted double power-law distributions to the data for roads and sewers, follow methods adapted from Clauset et al [32] and Corral and Deluca [20, 21]. The $k_{min}$ used for determining the break point between the two distributions and the fit of the lower distribution ($\gamma_{x2}$) were found using the method and R code provided by Clauset et al [20], while the fit for the upper "trunk" distribution ($\gamma_{x2}$) was estimated by MLE and the comparison of CDFs by the Kolmogorov-Smirnoff test (both implemented using Matlab R2016b) as suggested by Corral and Deluca for fitting discrete power-law distributions [21]. In all cases the upper distribution was calculated as a power-law distribution from k=3 to the

break point determined by the Clauset et al method. If for a given subnet there was no significant power-law tail, as determined by the Clauset et al. method, a single power law distribution was estimated from k = 3 to the highest node-degree for the subnet.

2. **Average Node-degree**: A measure of the average connectivity of each node in a network and calculated as $<k> = 2E/N$. The average node-degree of a network serves as an indicator of the types of connections that are present in a network and can help distinguish between a network characterized by a tree-like structure ($<k> = 2$) or a more grid like or cyclic structure ($<k> = 4$) [23, 24].

3. **Average Clustering Coefficient**: a measure based on the number of triplets within a network indicating the overall connectivity of neighboring nodes, and calculated as the number of closed triplets divided by the number of connected triplets of nodes. The clustering coefficient provides insight to the modularity or small world properties of a network. A high clustering coefficient would indicate that sections of the network (modules) are well connected within, but have only a few connections between different modules [29].

4. **Network Density**: A measure of the ratio of the number edges to the maximum possible number of edges, and calculated as the binomial coefficient (N/2). This measure has been used to assess the variability in network connectivity at various distances from the city center [45] and serves as indicator of connectivity at different spatial scales.

Table-1: Summary of study area road and sewer network features (Total Length, Primal Nodes and Edges, Dual Nodes and Edges, γ and Average Node-Degree)

|  | Total Length (km) | Edges (Primal) | Nodes (Primal) | Edges (Dual) | Nodes (Dual) | $\gamma_{x1}$ | $\gamma_{x2}$ |
|---|---|---|---|---|---|---|---|
| **Roads** | 8,790 | 67,989 | 69,866 | 44,044 | 24,900 | 2.5 | 3.1 |
| **Sewer** | 5,436 | 69,917 | 69,866 | 22,481 | 20,183 | 2.9 | 4.1 |

Table 2: Summary of topological metrics for individual road (R) and sewer subnets (S) (All values reported as mean ± σ). (*) indicates subnets rejected the null hypothesis of a power-law distribution for $\gamma_{x2}$

| Buffer Radius (km²) | $\gamma_{x1}$ | | $\gamma_{x2}$ | | Avg. Node-Degree | | Density | | Clustering Coefficient | |
|---|---|---|---|---|---|---|---|---|---|---|
| | R | S | R | S | R | S | R | S | R | S |
| 1.25 | 2.2 ± .6 | 3.3 ± .7 | * | * | 3.1 ± .7 | 2.1 ± .2 | .02 ± .006 | .01 ± .002 | .12 ± .06 | .16 ± .05 |
| 2.5 | 2.3 ± .3 | 3.3 ± .3 | 3.5 ± .5 | * | 3.3 ± .6 | 2.2 ± .1 | .006 ± .001 | .003 ± .001 | .12 ± .03 | .17 ± .04 |
| 5.0 | 2.3 ± .2 | 3.3 ± .2 | 3.4 ± .4 | * | 3.4 ± .5 | 2.3 ± .1 | .002 ± .000 | .001 ± .000 | .12 ± .02 | .18 ± .03 |
| 7.5 | 2.3 ± .2 | 3.3 ± .1 | 3.3 ± .2 | 7.4 ± 1.3 | 3.5 ± .4 | 2.3 ± .1 | .000 ± .000 | .000 ± .000 | .12 ± .02 | .18 ± .02 |

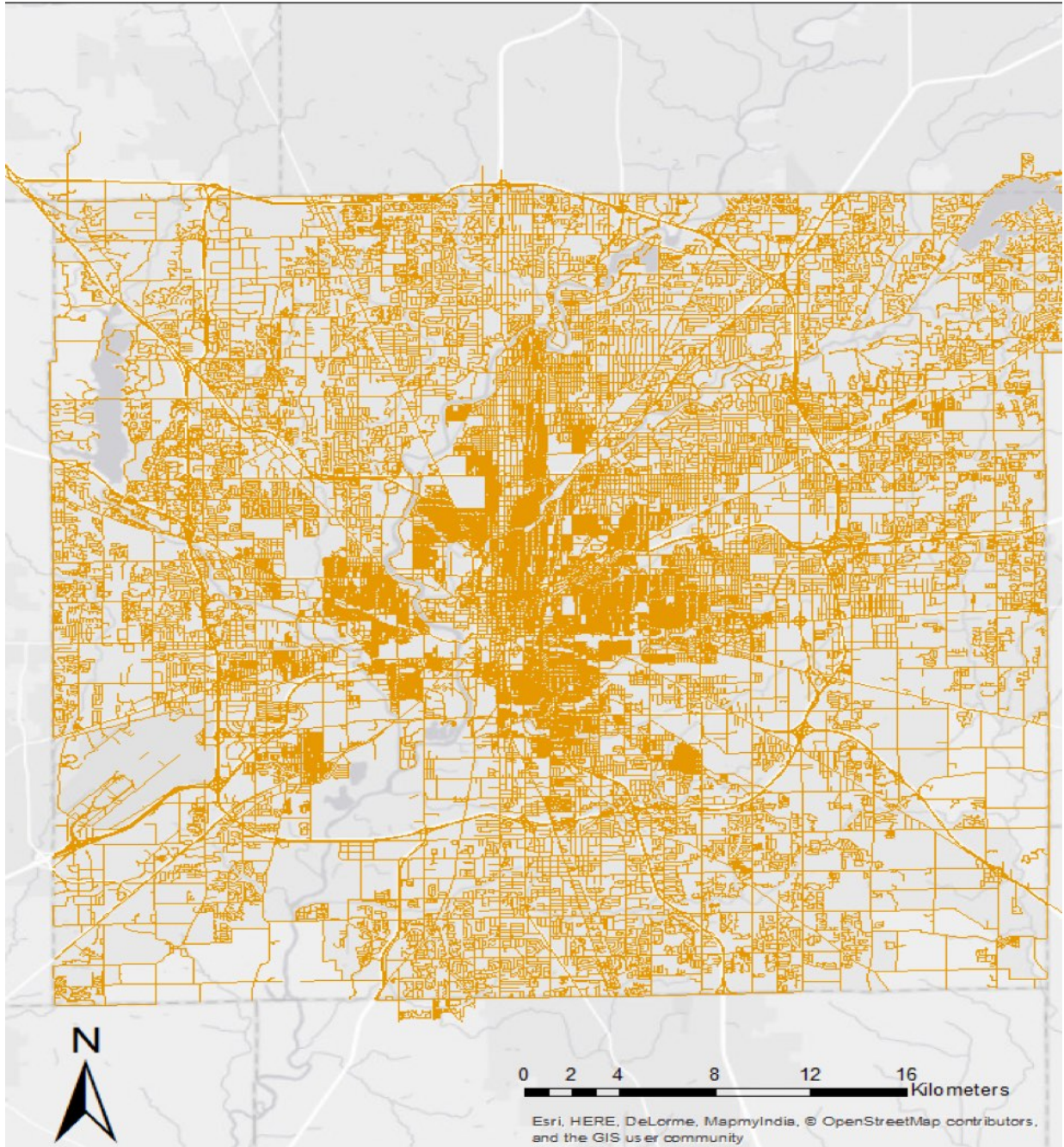

*Figure 1: Map of Midwest USA case study area. OpenStreetMap cartopgraphy map tiles are licensed under the Creative Commons license © OpenStreetMap contributors (CC BY-SA 2.0). Licensing information is available at http://www.openstreetmap.org/copyright*

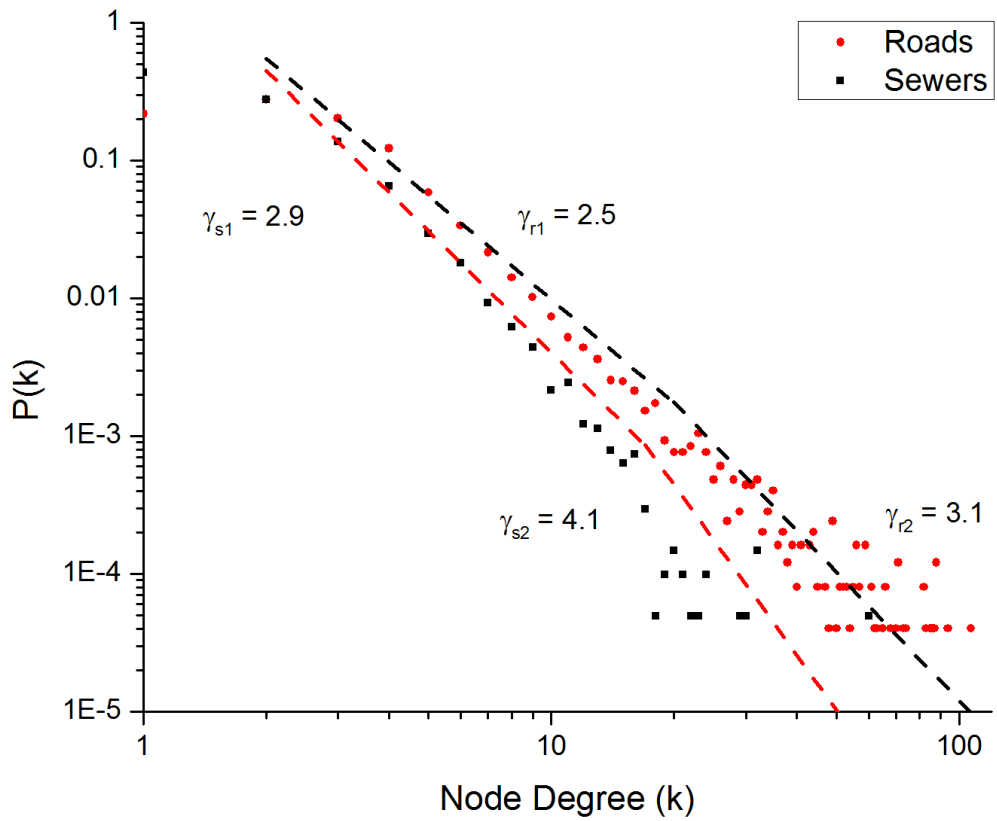

*Figure 2: Node degree distribution of dual mapped road and sewer networks*

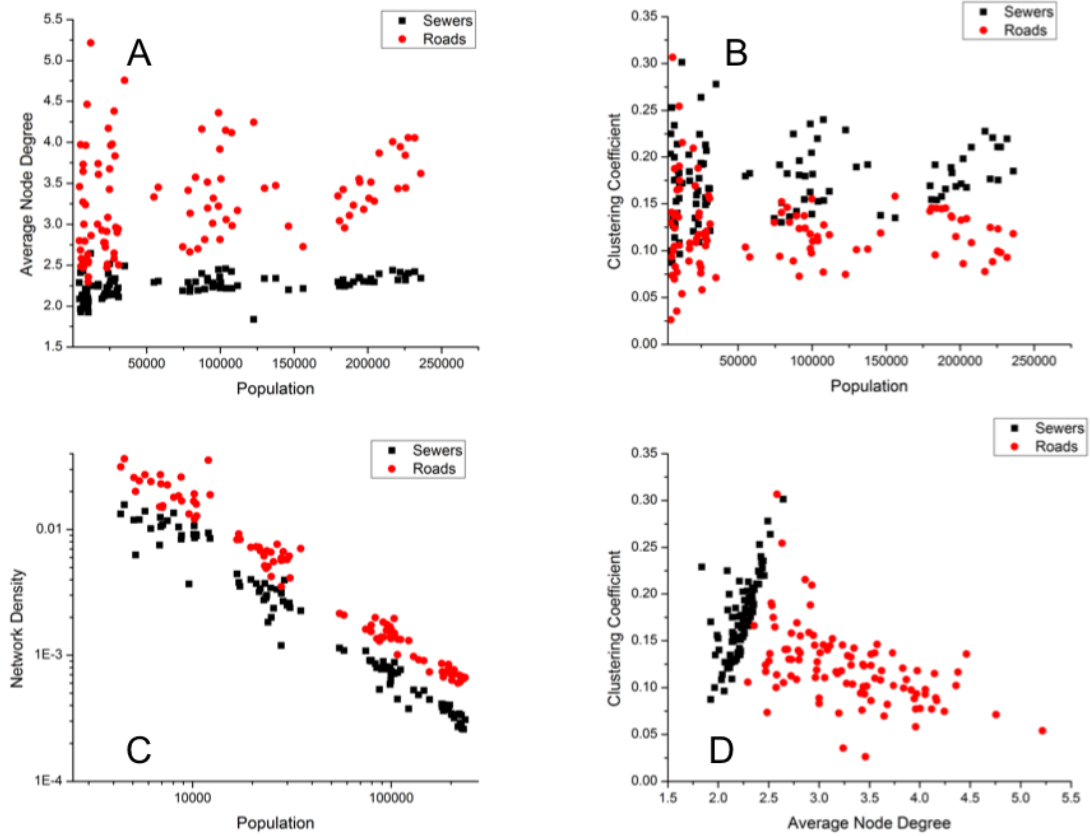

*Figure 3: Topological metrics generated for subnets (n=100) A) Average node degree B) Clustering coefficient C) Network Density D) Average node degree vs clustering coefficient*

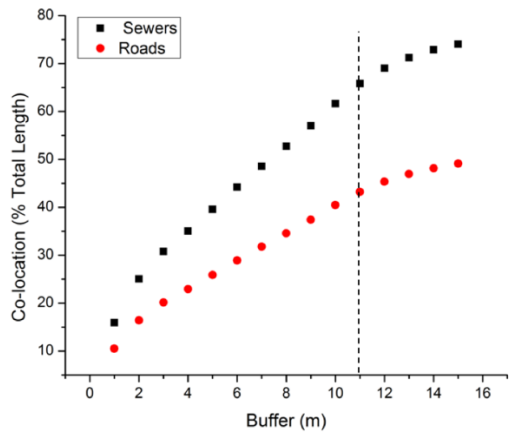 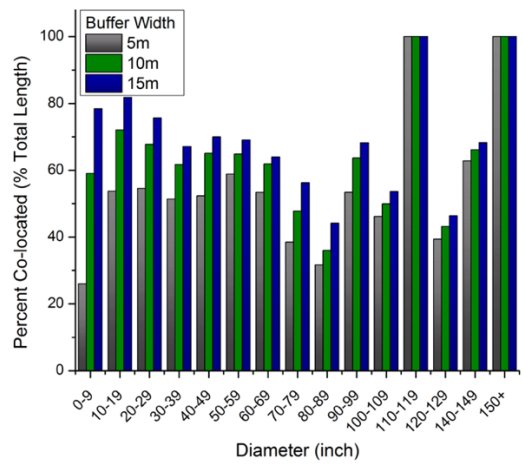

Figure 4: A) Co-location of roads and sewers at 1-15m buffers (Dashed line indicates mean width of right of way) B) Distribution of co-located pipe diameters

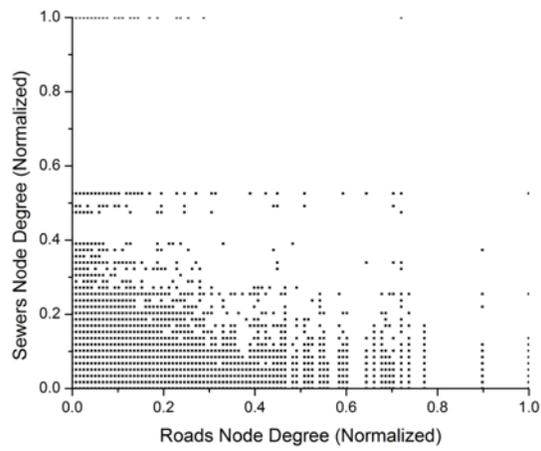 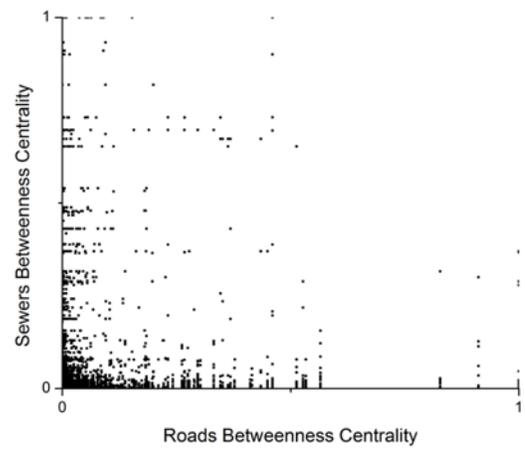

*Figure-5: A) Node degree of co-located roads and sewer pipes (Normalized 0 to 1) B) Betweenness Centrality of co-located roads and sewer pipes*

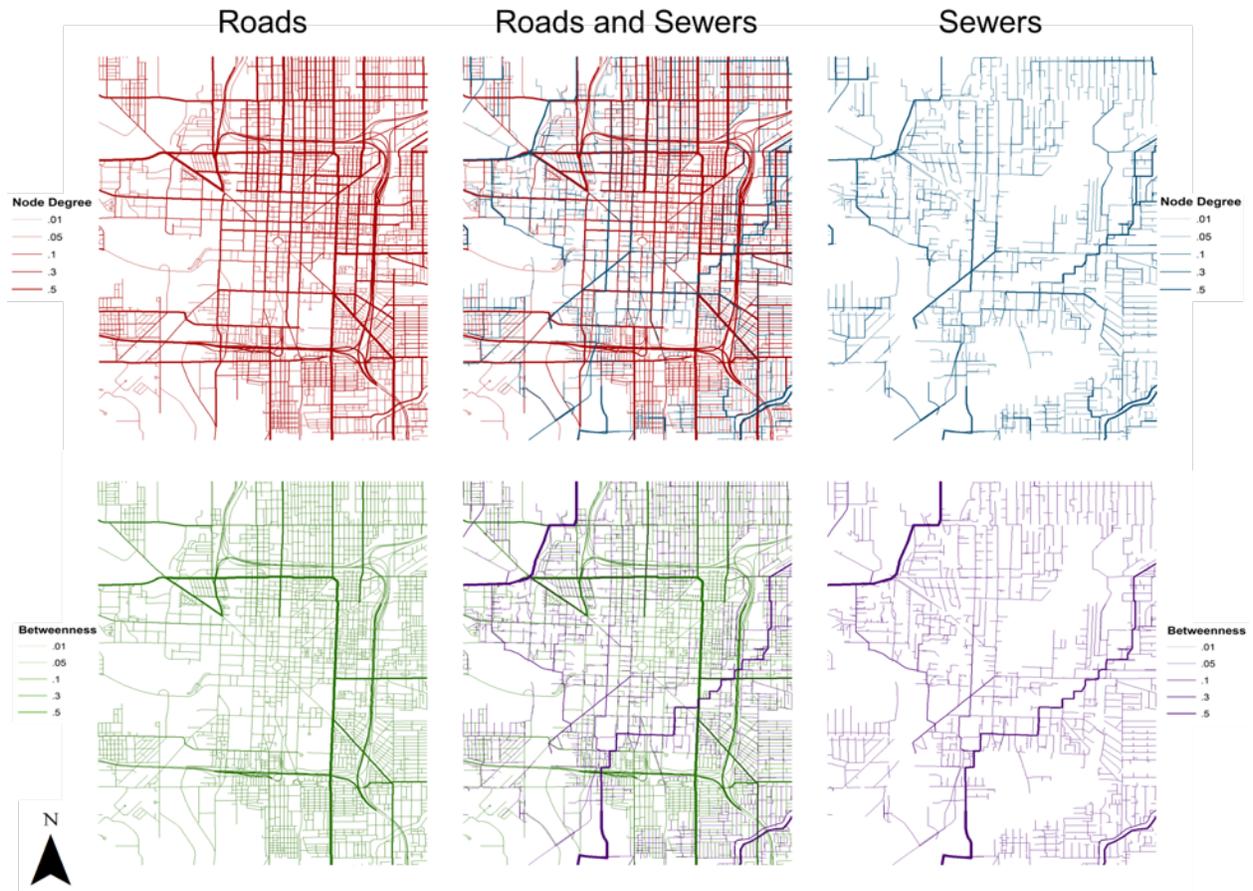

*Figure 6: Maps depicting the location of sanitary sewers and roads by node degree (normalized 0 to 1) and betweenness centrality. Line thickness represents magnitude of value. All maps in this figure were produced using ESRI ArcGIS ArcMap 10.1 http://www.esri.com/software/arcgis/arcgis-for-desktop. GIS data were provided by the city municipal GIS office and analyzed by the authors.*